\documentclass[pre,twocolumn,superscriptaddress,floatfix,showpacs,showkeys]{revtex4}

\usepackage{graphicx}
\usepackage[latin2]{inputenc}
\usepackage{amsmath}
\usepackage{epsfig}

\newcommand{\htop}{h_\text{top}}

\begin{document}

\title{Shear band formation in granular media as a variational problem}

\date{\today; version 1.0}

\author{T. Unger}
\affiliation{Department of Theoretical Physics, Budapest University of 
  Technology and Economics, H-1111 Budapest, Hungary}
\affiliation{Institute of Physics, University Duisburg Essen,
  D-47048 Duisburg, Germany}
\author{J. T\"or\"ok}
\affiliation{Department of Theoretical Physics, Budapest University of 
  Technology and Economics, H-1111 Budapest, Hungary}
\affiliation{Department of Chemical Information Technology, Budapest
  University of  
  Technology and Economics, H-1111 Budapest, Hungary}
\author{J. Kert\'esz}
\affiliation{Department of Theoretical Physics, Budapest University of 
  Technology and Economics, H-1111 Budapest, Hungary}
\author{D.E. Wolf}
\affiliation{Institute of Physics, University Duisburg Essen,
  D-47048 Duisburg, Germany}

\begin{abstract}
  Strain in sheared dense granular material is often localized
  in a narrow region called shear band. Recent experiments in a
  modified Couette cell \cite{Fenistein2003a,Fenistein2003b} provided
  localized shear flow in the bulk 
  away from the confining 
  walls. The non-trivial shape of the shear band  was
  measured as the function of the cell
  geometry. First we present a geometric argument for narrow shear
  bands which connects the function of their surface position with the
  shape in the bulk.
  Assuming a simple dissipation mechanism we show that the principle of
  minimum dissipation of energy provides a good description
  of the shape function. Furthermore, we discuss the possibility and
  behavior of shear bands which are detached from the free surface and are
  entirely covered in the bulk.
\end{abstract}



\maketitle

Granular media constitute an interesting field of research from the
point of view of both basic science and application. The intrinsically
nonlinear and dissipative nature of the interaction between the particles
leads to a great deal of interesting
phenomena including force networks, different kinds of instabilities,
clustering, complex flow properties \cite{DryGranu98,Jaeger96}. 
One of the most apparent instabilities occurring in granular media is
the formation of shear bands: At slow shear rate
strain is not distributed throughout the sample but appears in a
localized fashion along a rather narrow interface between two
essentially unstrained parts. Shear-banding was the subject of various
experimental and theoretical studies in the last few years
\cite{DryGranu98,Mueth00,Hartley03,Thompson91,Torok00,Schwedes03,Veje99}
and still presents significant difficulties for theoretical descriptions.

\begin{figure}[t]
\centerline{\epsfig{figure=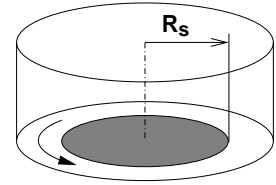,width=0.4\linewidth}}
\caption{Setup geometry. The rotating (white) and the stationary (gray)
 parts induce shear flow in the granular material held by the container.}
\label{fig:ExpGeom}
\end{figure}

Recently universal geometrical properties of shear bands
were discovered in a Couette geometry modified such that shear localization
near 
boundaries was avoided \cite{Fenistein2003a,Fenistein2003b}. 
The experimental setup was a cylindrical container filled with
grains up to height $H$. The bottom was split  into an outer ring
rotating with the container wall, and a stationary disk of radius $R_s$ 
in the center (Fig.~\ref{fig:ExpGeom})\footnote{The experiments
  \cite{Fenistein2003a,Fenistein2003b} were performed both with
  ``disk'' geometry 
  %described here 
  and with a split-bottomed Couette
  geometry, where an additional stationary inner cylinder is present.
  %with radius smaller then $R_s$. 
  In this paper only the ``disk'' geometry is discussed.}. 
Thus the outer and the inner part
of the material were rotated relative  to each other
which created a shear band with cylindrical symmetry:
It started at the perimeter of the stationary bottom disk
and extended through the bulk up to the free surface.
On the surface the angular velocity 
of the granular material as the function of the radius was measured. It 
follows to high accuracy an error function characterized by two 
parameters: the width $W$ and the center position $R_c$ of the shear zone.
The width grows with increasing height $H$ while the radius $R_c$ gets
smaller. Interestingly, the surface position  $R_c$ proved to be very
robust: It depends only on two length parameters $H$ and
$R_s$, but not on the particle properties 
%(size, shape, friction, hardness),
nor on the shear rate. This is contrary to the
width of the band, which is affected by the size and shape of the
grains but is insensitive to changes of the slip radius $R_s$. The fact
that $R_c$ and $W$ depend on different control parameters suggests that the
two quantities can be studied separately.

In this Letter we address the problem of the position of the shear band.
By suitable choice e.g. of the particle size the width can be made
arbitrarily small compared to $H$ and $R_s$. This justifies to model the
shear band as an infinitely thin layer
that represents  the boundary 
between two blocks of material within which no flow occurs. We derive
the shape of the band from a \emph{variational principle}. Note that
optimization 
principle was applied already to shear band 
formation in a different context \cite{Torok00}.

%We derive the shape 
%of the shear band from a variational principle.
%We assume that the mechanism responsible for the position of the shear
%band can be studied separately from its width or flow profile.
%Therefore we model the shear zone as an infinitely thin shear band
%that represents  the boundary 
%between two blocks of material within which no flow occurs.
%We note that the optimization principle was applied already to shear band
%formation in a different context \cite{Torok00}. 

\begin{figure}
\centerline{\epsfig{figure=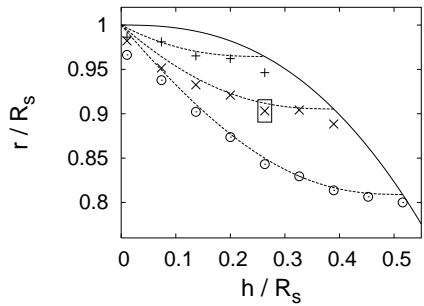,width=0.7\linewidth}}
\caption{
The symbols are experimental data showing the shear zone radius $r$
  measured in the bulk at height $h$, taken from \cite{Fenistein2003b}.
  $+$, $\times$ and $\circ$ correspond filling heights $25$, $37$ and
  $49$ mm respectively, $R_s$ is $95$ mm.
  The rectangle in 
  the middle  shows the estimated errors in both directions for all data
  \cite{MvHpriv} (plotted only for one data point).
  The solid line is the experimentally found fit curve
  for the surface positions.
  The dashed lines are the calculated bulk-positions based on the
  solid curve.
}
\label{fig:BulkSurfRel}
\end{figure}

\textbf{Surface - bulk relation.} In the experiments the position
$R_c(R_s,H)$ on the free surface was found \cite{Fenistein2003b} 
to scale like
\begin{equation}
R_c(R_s, H)= R_s (1-(H/R_s)^{\alpha}) 
\label{Rc5per2}
\end{equation}
with $\alpha \approx 2.5$ for the experimentally accessible values of
$H/R_s$.
It is much more difficult to measure the position of the shear band in the
bulk, $r(h)$, for fixed $R_s$ and $H$.  
Nonetheless, the experimental data clearly show that the
bulk profile follows another form than $R_c(R_s, H)$, and 
the bulk radius at height $h$ depends also on the filling height $H$
(Fig.~\ref{fig:BulkSurfRel}).

We show that \emph{the bulk profile is determined once
the surface positions $R_c(R_s,H)$ are given}. Let us take a system with total
height $H$ and find the position of
the shear band $r$ at height $h<H$. The subsystem above $h$ can be
regarded as a smaller system with height $(H-h)$ and with slip
radius $r$ at the bottom. Pressure and boundary conditions are the
same, and the difference in the width is neglected in the narrow band
approximation.
% From the position $R_c$ of the shear band on the
%surface one cannot decide, whether it belongs to a small system with 
%slip radius $r$ and filling height $H-h$ or to a large system with
%$R_s$ and $H$.
We conclude that 
\begin{equation}
R_c(R_s, H) = R_c(r,H-h)\, .
\label{surface-bulk-relation}
\end{equation}
Knowing the function $R_c(R_s,H)$ thus allows to calculate the shape
$r(h)$ of the shear band throughout the whole system. Note, if the 
simple size scaling holds for the surface radius that $R_c/R_s$ depends
only on $H/R_s$ (as it was found in \cite{Fenistein2003b}) then it follows
immediately also 
for the bulk function due to Eq.~(\ref{surface-bulk-relation}):
\begin{equation}
\frac{r}{R_s} = f\left( \frac{H}{R_s},\frac{h}{R_s}\right) \ .
\label{bulk-scaling}
\end{equation}
Based on Eq.~(\ref{surface-bulk-relation}) an 
 explicit functional form of the bulk profile can be obtained using the
 experimental fit function of $R_c$:
\begin{equation}
h = H - r \left[1-\frac{R_s}{r} (1-(H/R_s)^{\alpha})\right]^{1/\alpha} \ .
\label{bulk-Rc5per2}
\end{equation}
The resulting curves for some filling heights are plotted in
Fig.~\ref{fig:BulkSurfRel} using $\alpha=2.5$ and the 
comparison with the experimental data shows very good agreement.

\textbf{Variational principle}. In order to describe the form of the shear
band our idea is to apply 
the \emph{principle of least dissipation} \cite{Onsager1931} (which is a
common treatment of time-independent irreversible 
  phenomena \cite{statphys}).
Therefore we require a steady state flow that matches the outer
constraints but provides the minimum rate of energy dissipation. 

Applying this to the cylindrical geometry within the narrow band
approximation 
 the question of the 
shape is traced back to  a variational problem among the functions
$r(h)$ (where $H$ and $R_s$ are kept fixed) with the condition
$r(0)=R_s$ while the other boundary at $H$ is free. The dissipation
rate is given by the 
sliding velocity between the two sides $r(h) \omega$ times the shear stress
$\sigma_{tn}$ integrated over the whole shear band. Up to a constant
factor, the expression to be minimized is  
\begin{equation}
 \int_0^H r^2 \sqrt{1+ \left( {dr}/{dh}\right)^2} \, \sigma_{tn} {d}h =
 \textrm{min.} 
\label{vari}
\end{equation}
This quantity represents not only the \emph{dissipation rate} but also the
\emph{mechanical torque} that the rotating and stationary part of the
system exert on each 
other. %
Therefore the \emph{least
dissipation} for this specific geometry is
equivalent to the \emph{minimal torque} which gives further 
justification of this 
approach: it is plausible that the yielding surface is established where
the resistance against the outer constraint is the smallest, i.e.\ where the
material is the weakest \cite{Torok00}.

\textbf{Sliding model}. For the shear stress in Eq.~\ref{vari}
% For the full description of the variational
%problem (Eq.~\ref{vari}) one needs the expression of the shear
%stress. 
%In order to be able to proceed 
we will use a very simple sliding
model. The shear stress in the yielding surface is taken similar to the
Coulomb friction between two solid bodies:
It acts against the sliding direction,
its magnitude is proportional to the normal pressure pressing the two
sides against each other, but it is independent of the sliding velocity. We
assume hydrostatic pressure
i.e.\ proportional to the depth \footnote{Here we ignore anisotropy
  effects.}. 
Thus Janssen effect is neglected, which is naturally justified if H is smaller
than the container width. In our dynamical situation, however, we expect
that the applicability of the hydrostatic pressure can be extended even for
larger filling heights: The shear band (due to many collisions and slip
events) acts as a source of small vibrations in the whole system and can
cause slight creep at the particle-wall contacts inhibiting the particles
to keep their original anchoring position. Finally they transmit their load
to the next particle below rather than to the side wall and therefore the
whole weight will be carried by the bottom.

This sliding model leads to the variational problem
%a well defined variational problem:
\begin{equation}
 \int_0^H r^2 \sqrt{1+ \left( {dr}/{dh}\right)^2} \, (H-h) {d}h =
 \textrm{min.} \ .
\label{sliding-vari}
\end{equation}
% where the resulting behavior can be analyzed.

The solutions which minimize the integral have automatically the scaling
property given by Eq.~(\ref{bulk-scaling}). The reason of this data
collapse is that taking a
$\lambda$  times larger system (i.e.\ taking $\lambda R_s$, $\lambda H$,
$\lambda r(h/\lambda)$ instead of $R_s$, $H$, $r(h)$) changes the value of
the integral only by a constant factor ($\lambda^4$) thus it represents the
same variational problem.

\textbf{Numerical results}.
The function $r(h)$ is discretized and the minimization
 is performed numerically  
based on genetic optimization: 
$r(h)$ is varied randomly but only
the changes that lower the value of the left hand side in
Eq.~(\ref{sliding-vari}) 
,i.e., reduce the dissipation, are admitted. During the optimization the
noise level is 
continuously decreased and the final state $r(h)$ is regarded as  one local
minimum of the variational
problem. The landscape where the minimum has to be found is simple (see later).

\begin{figure}
\centerline{\epsfig{figure=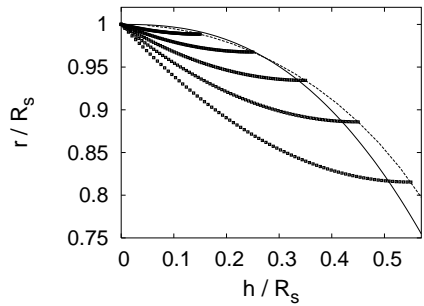,width=0.7\linewidth}}
\caption{Results obtained from the variational principle.
  Symbols show bulk profiles, from top to bottom $H/R_s=0.15$, $0.25$,
  $0.35$, $0.45$, $0.55$ respectively.
  The two lines denote the surface positions as the function of
  the total height. The dashed line comes from our model, the solid line is
  the experimental fit function.}
\label{fig:NumericShape}
\end{figure}

Results shown in Fig.~(\ref{fig:NumericShape}) reproduce nicely the
qualitative behavior found in the experiment: the concave shear bands
appear in the bulk and build up a convex confining shape
of the surface positions as the filling height is varied. The shear radii at a
fixed bulk height $h$ and 
also at the top get smaller with increasing filling height. 

The realistic bulk profiles provided by the principle of minimum dissipation
can be interpreted easily. The benefit gained by having such a curved
cylindrical 
shape (slimmer at the top) in place of a regular cylinder is twofold:
First, the surface of
the shear band can be reduced by letting the regular cylinder deform
and by pulling the shape towards the center a bit at a fixed bottom radius,
similarly to a soap membrane spanned between a ring and a plate
where the plane of the ring is parallel to the plate.
Second, smaller radius results in smaller sliding velocity (or thinking of
minimizing the torque it represents smaller lever).
Therefore
one can roughly think of the bulk profiles shown in
Fig.~(\ref{fig:NumericShape}) 
as equilibrium situations where the reduction
of the sliding velocity is counterbalanced by the increase in the surface
due to going beyond the minimum surface.

The quantitative agreement with the
experimental fit function (Eq.~\ref{Rc5per2}) is also surprisingly good
given the crude assumptions we made and the fact that our model
contains no free fit parameters. The difference of the theoretical and
the experimental values of $R_c$ is less than $20$~\% of $R_s-R_c$.

We can more easily analyze the limit $H/R_s
\rightarrow 0$ than in the experiment, where $R_s$-values are
limited by the container size and a power law fit to the data at low
filling heights is not very precise. 
From our variational principle we find an exponent $\alpha = 2$ for
small $H/R_s$. In the region $0.4 < H/R_s < 0.7$ there is no clean power
law any more and the effective exponent increases.
In our model this seems to be due to a phase transition at about
$H/R_s \approx 0.7$ to a different shape of the shear band, which we
discuss next. 

\begin{figure}
\centerline{\epsfig{figure=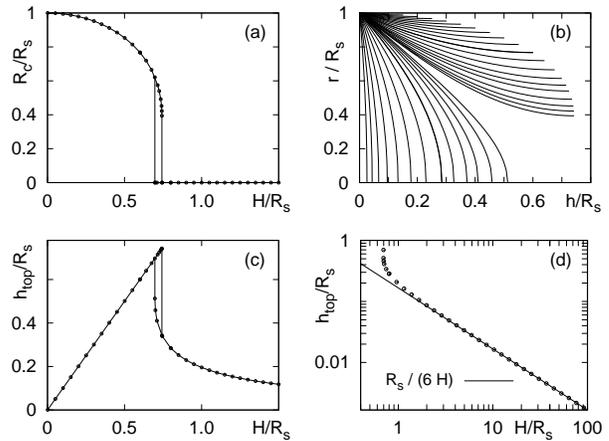,width=1.0\linewidth}}
\caption{Open and closed shear bands with cylindrical symmetry.
  (a) upper radius, (b) bulk profiles for several filling heights ($0< H/R_s
  < 7$), 
  (c,d) height of the shear band $\htop$.}
\label{fig:pro-4fig}
\end{figure}

What is the predicted behavior of the shear band for filling heights larger
then those reported in \cite{Fenistein2003a,Fenistein2003b} ? For large
enough $H/R_s$ the class of the ``open'' solutions discussed so far is replaced
by a new type of solutions: The shear band, instead of running up to the free
surface and having a circle on the top as its upper edge, closes forming a
cupola-like shape (still with bottom radius $R_s$). In that case the
material covered by the  ``closed'' shear band is at rest while the
material around and above  (including the whole free surface) is rotating.
Several ``open'' and ``closed'' profiles can be seen in
Fig.~(\ref{fig:pro-4fig}b) obtained for various values of
$H$. Fig.~(\ref{fig:pro-4fig}a) shows the upper radius of the 
shear bands which is characteristic of the ``open'' solutions but becomes
zero for the ``closed'' ones. For these cupola shapes 
  \footnote{The ``closed'' profiles appear in the integral of
  Eq.~(\ref{sliding-vari}) in the way that the function $r(h)$ becomes zero
  for $\htop\le h \le H$. Therefore this region has no contribution to the
  dissipation and practically the integral is meant over the interval
  $[0,\htop]$.}
a more relevant parameter is their 
heights $\htop \le H$ in the center, plotted in 
Fig.~(\ref{fig:pro-4fig}c). For ``open'' profiles $\htop$ equals
simply the system height.

Interesting is the behavior of the parameter $\htop$. It is a monotonically
decreasing function of the filling height once it is detached from
$H$, i.e.\ large filling heights press the shear band to the
bottom. Solving the Euler-Lagrange equation for the variational
problem neglecting terms of higher than first
order in $dh/dr$ and taking the limit of $H\gg \htop$ one obtains
\begin{equation}
h(r) = \htop -\frac{1}{6H} r^2\, , \qquad \htop = \frac{1}{6H} R_s^2
\, .
\end{equation}
Fig.~(\ref{fig:pro-4fig}d) shows the numerical solution of
$\htop$ : it is in excellent agreement with the approximate analytical
solution. In the limit $H/R_s\rightarrow\infty$ the material forms
one solid block and the sliding occurs on the surface of the bottom disk.

The $H$ dependence of $\htop$ has the following reason:
Larger $h$ corresponds
to lower pressure and thus to smaller shear-resistance;
therefore it is worth to raise the shear band into a cupola-shape
even if its surface becomes larger than the disk at the bottom. 
Note, however, that stronger gravity would not affect the shape: It
gives a larger pressure gradient but 
this constant factor has no impact on the variational problem. 
By contrast, increasing $H$ (or, alternatively, applying additional
pressure at the surface) makes the {\em relative} pressure change
smaller near the bottom  which results in weaker uplifting
``force''. This is why the cupola height $\htop$ approaches zero 
for large $H$ (respectively large pressure).

\begin{figure}
\centerline{\epsfig{figure=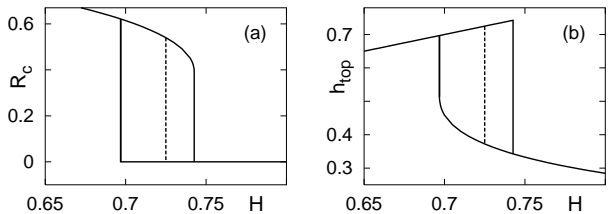,width=1.0\linewidth}}
\caption{Hysteresis emerging in the variational problem. Upper radius (a) and
  height (b) of the shear band in units of $R_s$. The dashed line denotes the
  transition of the global minimum.
}
\label{fig:hyster}
\end{figure}

At intermediate $H$ (around $0.7 R_s$)
there exists
a small region of filling heights where the variational problem
(Eq.~\ref{sliding-vari}) exhibits two local minima. The behavior of
the local minima corresponds to a first order 
phase transition, where the two phases are the two types of shear band
profile (Fig~\ref{fig:hyster}). Following the ``closed'' solution as $H$ is
decreased 
its height grows till the top of the cupola reaches the
surface at system height $H_1$. At this point the ``closed'' solution
becomes unstable, $R_c$ at
the top runs suddenly to a non-zero value (Fig.~\ref{fig:hyster}a). Regarding
the other 
direction (increasing $H$) the upper edge of the
``open'' shape is pulled towards the center then at height $H_2$ it shrinks
to one point which is followed by a jump into a ``closed'' shape
(discontinuity in 
$\htop$ see in Fig.~\ref{fig:hyster}b). The interval $[H_1,H_2]$ defines
the region 
where two local minima exist, outside this region there is only one phase
possible. Local minima are physically meaningful in the presence of a kinetic
barrier. The latter can be provided by the difference between the static
and dynamic friction coefficients, usually present in granular systems.

In this Letter we have presented a theoretical analysis of recent experiments
[1,2] on shear band formation in granular media. We used
the approximation of narrow shear bands. First we showed a geometric
argument which related the position of the surface endpoints of the
band to the bulk shape. We calculated the shape of shear bands
from a variational principle and the results are in good agreement
with the experiments. The theory provides with a number of
predictions. i) There is a transition in the shape of the shear band
as a function of the filling height $H$: For low values of $H$ the
shear band is an open curved cylinder which ends on the surface while
for large $H$ a cupola is formed. ii) This transition is of the first
order, accompanied by a hysteresis. iii) The height of the cupola is
proportional to $R_s/H$. These predictions should be
experimentally accessible.

An interesting question we did not address in this Letter is the
nucleation kinetics at the first order phase transition from open to
closed shapes. It is conceivable that the finite width of the
shear bands plays a role here. Further progress in this direction
requires a continuum theory going beyond the narrow band approximation
we used here.

We would like to thank to D.\ Fenistein and M.\ van Hecke for useful
discussions. Partial support by grant OTKA T035028 and by the
German-Hungarian Cooperation Fund is acknowledged.


\begin{thebibliography}{99}

\bibitem{Fenistein2003a} D. Fenistein and M. van Hecke, Nature \textbf{425}, 256 (2003).

\bibitem{Fenistein2003b} D. Fenistein, J. W. van de Meent and M. van Hecke,
  cond-mat/0310409 (2003).

\bibitem{DryGranu98} \emph{Physics of Dry Granular Media},
 eds. H. J. Herrmann, J.-P. Hovi,  S. Luding (Kluwer Academic Publishers,
 Dordrecht, 1998).

\bibitem{Jaeger96} H. M. Jaeger, S. R. Nagel, R. P. Behringer,
%
Rev. Mod. Phys. \textbf{68}, 1259 (1996)

\bibitem{Mueth00} D. M. Mueth, G. F. Debregeas, G. S. Karczmar, P. J. Eng,
  S. R. Nagel and H. M. Jaeger,
%
Nature \textbf{406}, 385 (2000).

\bibitem{Hartley03} R. R. Hartley and R. P. Behringer,
%
%
Nature \textbf{421}, 928 (2003). 

\bibitem{Thompson91} P. A. Thompson and G. S. Grest,
%
Phys. Rev. Lett. \textbf{67}, 1751 (1991).

\bibitem{Torok00} J. T\"or\"ok, S. Krishnamurthy, J. Kert\'esz and
S. Roux,
%
%
Phys. Rev. Lett. \textbf{84}, 3851 (2000).

\bibitem{Schwedes03} J. Schwedes,
Gran. Mat. \textbf{5}, 1 (2003).

\bibitem{Veje99} C. Veje, D. Howell, and R. P. Behringer, 
Phys. Rev. Lett \textbf{82}, 5241(1999).


\bibitem{MvHpriv} M. van Hecke, private communication.

\bibitem{Onsager1931} L. Onsager,
Phys. Rev. \textbf{37}, 405 (1931), Phys. Rev. \textbf{38} 2265 (1931).

\bibitem{statphys} K. E. Reichl: A Modern Course in Statistical Physics 2nd
  ed. 1998. John Wiley, New York.

\end{thebibliography}
\end{document}